\documentclass[12pt]{article}
\usepackage{a4wide,epsf,epsfig}

\newcommand{\phitwopi}{\Phi_{2\pi}}

\begin{document}
\begin{flushright}
WU B 99-30 \\
SLAC-PUB-8320 \\
PITHA 99/39\\
hep-ph/9912364
\end{flushright}

\begin{center}
\vskip 3.5\baselineskip
{\Large\bf The perturbative limit of the \\[1ex] 
two-pion light-cone distribution amplitude} 
\vskip 2.5\baselineskip
M. Diehl$^{1}$, Th.~Feldmann$^{2}$, P. Kroll$^{3}$, C. Vogt$^{3}$
\vskip \baselineskip
{}$^{1}$ SLAC, P.O.~Box 4349, Stanford, CA 94309, U.S.A.
\vskip \baselineskip
{}$^{2}$ Institut f\"ur Theoretische Physik~E, RWTH Aachen, 52056
         Aachen, Germany
\vskip \baselineskip
{}$^{3}$ Fachbereich Physik, Universit\"at Wuppertal, 42097 Wuppertal,
         Germany 
\vskip \baselineskip

\vskip 2.5\baselineskip

\textbf{Abstract} \\[0.5\baselineskip]

\parbox{0.9\textwidth}{We consider pion pair production in two-photon
collisions, $\gamma^*\gamma\to\pi^+\pi^-$, at large photon virtuality
$Q^2$ and center-of-mass energy $\sqrt s$ in the hard-scattering
picture. In the limit where $s$ is large but $s \ll Q^2$ we derive an
expression for the two-pion light-cone distribution in terms of the
conventional pion distribution amplitudes. Our result reproduces the
well-known scaling behavior, helicity selection rule and symmetry
relations in this regime.}

\vskip 1.5\baselineskip
\end{center}

%\baselineskip1.8em

% end of title page

{\bf 1.}  The two-pion distribution amplitude has been introduced in
connection with pion pair production $\gamma^*\gamma\to \pi^+\pi^-$ in
the limit of large $Q^2$ and small $s$ \cite{Mueller,Diehl1}. This
process is related by crossing to deeply virtual Compton scattering at
large $Q^2$ and small $t$, which is known to factorize into a hard
photon-parton scattering and a skewed parton distribution
\cite{Mueller,Ji,Radyushkin}. Hadron pair production factorizes in a
completely analogous fashion into a short-distance dominated
subprocess $\gamma^*\gamma\to q\bar{q}$ or $\gamma^*\gamma\to gg$ and
a generalized distribution amplitude describing the soft hadronization
of partons into two hadrons \cite{Freund}. In Refs.~\cite{Diehl1} and
\cite{Polyakov1} general properties of the two-pion distribution
amplitude have been discussed. For small $s$ it has been calculated
from the instanton model of the QCD vacuum \cite{Polyakov2}.

On the other hand, in the kinematical region of large $s,\,-t,\,-u$,
two photon annihilation into pion pairs is known to factorize into two
single-pion distribution amplitudes and a hard-scattering amplitude to
be calculated perturbatively \cite{BrodskyLepage}. The case of two
real photons has been treated in Ref.~\cite{BrodskyLepage2}. The
authors of Ref.~\cite{Gunion} have discussed the general situation
where both photons are virtual and investigated the applicability of
perturbative QCD to exclusive large-angle processes.

We are going to show that in the limit $\Lambda^2\ll s\ll Q^2$, with
$\Lambda$ being a typical hadronic scale of the order of 1 GeV, both
factorization schemes merge. This allows us to represent the two-pion
distribution amplitude of \cite{Diehl1} in terms of the two ordinary
single-pion distribution amplitudes as defined in
\cite{BrodskyLepage}. We will show that the general properties of the
two-pion distribution amplitude are satisfied in this representation,
and discuss factorization properties, scaling behavior and helicity
selection rules of the process.

The outline of the paper is as follows. First we will introduce our
notation, define the two-pion distribution amplitude and discuss its
properties. Then we show how to obtain the perturbative limit of the
two-pion distribution amplitude from the hard-scattering picture and
reproduce its general properties. We conclude with our summary. \\

\begin{figure}[ht]
\begin{center}
\psfig{file=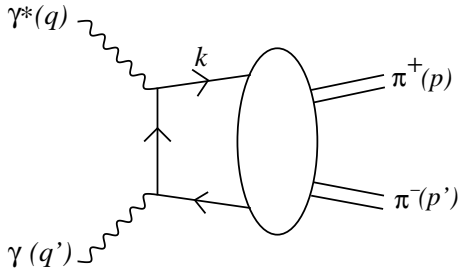,bb=20 220 580 610,width=7cm,angle=0}
\caption{Factorization of the process $\gamma^*\gamma\to\pi^+\pi^-$
for $Q^2 \gg s$. The hard scattering subprocess is shown at leading
order in $\alpha_s$, and the blob represents the two-pion distribution
amplitude. The second contributing graph is obtained by interchanging
the photon vertices.} \label{sketch}
\end{center}
\end{figure}

{\bf 2.} We denote the momenta of the virtual and real photon by $q$
and $q'$, respectively, and define $P=p+p'$, where $p$ and $p'$ are
the pion momenta (for notation see also Fig.~\ref{sketch}). Moreover,
we use the quantities $Q^2=-q^2$ and $s=P^2$ as usual. For our purpose
it is advantageous to choose a frame in which both the virtual photon
and the produced partons are moving rapidly in the positive
$z$-direction. We work in the Breit frame, where after introducing the
lightlike vectors $v=(1,0,0,1)/\sqrt{2}$ and $v'=(1,0,0,-1)/\sqrt{2}$
we can write
\begin{eqnarray}
q&=&\frac{Q}{\sqrt{2}}\, (v-v'), \quad 
q'=\frac{Q^2+s}{\sqrt{2}Q}\, v'.
\end{eqnarray}
The light-cone fraction of the quark with momentum $k$ is defined by
\begin{eqnarray}
 z=\frac{k^+}{P^+}, \label{def z}
\end{eqnarray}
and we introduce a parameter $\zeta$, which describes how the produced
pions share the total light-cone momentum $P^+$, in the following way:
\begin{eqnarray}
\zeta=\frac{p^+}{P^+}. \label{def zeta}
\end{eqnarray} 
The pion momenta can be written as  
\begin{eqnarray}
 p=\frac{Q}{\sqrt{2}}\,\zeta v
  +\frac{s}{\sqrt{2}Q}\,(1-\zeta) v'+p_\perp ,\quad  
 p'=\frac{Q}{\sqrt{2}}\,(1-\zeta) v
   +\frac{s}{\sqrt{2}Q}\,\zeta v'-p_\perp , 
\end{eqnarray}
where $p_\perp$ is transverse to $v$ and $v'$. The on-shell condition 
immediately provides a kinematical constraint for $\zeta$:
\begin{eqnarray}
 \zeta(1-\zeta)=\frac{m_\pi^2+{\bf
 p}_\perp^2}{s}\ge\frac{m_\pi^2}{s}. \label{zeta constraint} 
\end{eqnarray}
Working in the limit of large $s$ we will henceforth neglect the pion
mass. In the particular frame we are using, the minus and transverse
components of $p$ and $p'$ are small for $s \ll Q^2$. The relation
between the longitudinal momentum fraction $\zeta$ and the momentum
transfer $t$ is given by
\begin{eqnarray} 
 \zeta=1+\frac{t}{Q^2+s}.
\end{eqnarray}
We then have, up to corrections of ${\cal O}(s/Q^2)$,
\begin{eqnarray} 
 t=-(1-\zeta)Q^2, \quad u=-\zeta Q^2. \label{tu}
\end{eqnarray}

Having introduced our notation and kinematics we now turn to a
discussion of the two-pion distribution amplitude $\phitwopi^q$ for a
quark flavor $q$. It is defined by the following matrix element of
quark field operators \cite{Diehl1}:
\begin{eqnarray}
 \phitwopi^q(z,\zeta,s)=\frac{1}{2\pi}\int_{-\infty}^{\infty} dx^-
      e^{-iz(P^+x^-)}\langle\pi^+(p)\pi^-(p')| \,
      \overline{\Psi}_q(x^-v')\gamma^+\Psi_q(0)\, |0\rangle, 
  \label{def twopida} 
\end{eqnarray}
where the integral is over the minus component of the spatial
separation of the fields. Here we are explicitly using light-cone
gauge $A^+=0$ for the gluon field, where the path-ordered exponential
of gauge fields, usually present in (\ref{def twopida}) in order to
give a gauge invariant expression, can be dropped. According to
\cite{Diehl1}, factorization of the process $\gamma^* \gamma \to \pi^+
\pi^-$ means that we can write its amplitude as a convolution of a
hard scattering and the two-pion distribution amplitude. To leading
order in $\alpha_s$ we have
\begin{eqnarray}
 {\cal M}_{\lambda\lambda'}(\zeta,s)= 
% -\frac{1}{2}\,
  \frac{1}{2}\,
  \delta_{\lambda\lambda'}
  \sum_q\, e^2 e_q^2\,\int_0^1 dz \, \frac{2z-1}{z\bar{z}} \,
  \phitwopi^q(z,\zeta,s), \label{helamp0} 
\end{eqnarray}
where here and in the following we use the generic notation
$\bar{a}\equiv 1-a$. The sum in (\ref{helamp0}) is over all quark
flavors. $e_q$ is the quark charge in units of the positron charge
$e$, and $\lambda$ and $\lambda'$ respectively denote the helicities
of the virtual and real photon in the Breit frame. The helicity
selection rule expressed through $\delta_{\lambda\lambda'}$ has its
physical origin in angular momentum conservation and the approximation
of massless quarks in the hard scattering.

Note that the leading order expression is completely independent of
the photon virtuality $Q$. The leading logarithmic corrections to
(\ref{helamp0}) result from the evolution of the two-pion distribution
amplitude. It is governed by the well-known ERBL evolution equations
\cite{BrodskyLepage,EfremovRadyushkin} in the factorization scale
$\mu^2$, which is naturally chosen to be of order $Q^2$ here. In
contrast to the ordinary distribution amplitudes the two-pion
distribution amplitude can in general be complex. Its imaginary part
is due to final state interactions between the two pions, and to the
formation of resonances and other intermediate hadronic states.
Invariance under charge conjugation provides the following symmetry
relation:
\begin{eqnarray}
 \phitwopi^q(z,\zeta,s)=-\phitwopi^q(\bar{z},\bar{\zeta},s) . 
\label{sym}
\end{eqnarray}
It was also remarked in~\cite{Diehl1} and elaborated on
in~\cite{Polyakov1} that the two-pion distribution amplitude may be
seen as the crossed version of skewed parton distributions, which
naturally appear in deeply virtual Compton scattering off pions.

A powerful constraint on the $\zeta$ dependence of the two-pion
distribution amplitude is that the $n$-th moment $\int_0^1 dz\,
(2z-1)^{n-1}\, \phitwopi^q(z,\zeta,s)$ is a polynomial of order $n$ in
$\zeta$ \cite{Polyakov1}. This is a consequence of Lorentz invariance
and completely analogous to the so-called polynomiality condition for
skewed parton distributions~\cite{JiRad}. For the first moment one has
in particular
\begin{equation}
  \label{sum-rule}
\sum_q e_q \int_0^1 dz\, \phitwopi^q(z,\zeta,s) = (2\zeta-1) F_\pi(s),
\end{equation}
where $F_\pi(s)$ is the timelike pion form factor. Our phase
conventions for pion states are such that
\begin{equation}
  \langle\pi^+(p)\pi^-(p')|\, j^\mu(0) \,|0\rangle = 
  e (p-p')^\mu\, F_\pi(s) ,
\end{equation}
where $j^\mu$ is the electromagnetic current. \\

\begin{figure}[ht]
\begin{center}
\psfig{file=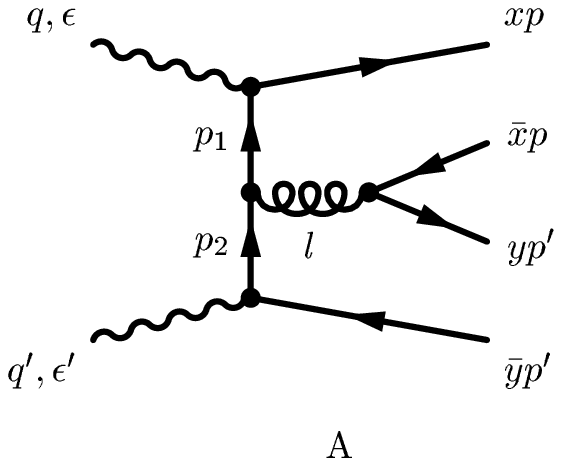,bb=200 640 380 790,width=5cm,angle=0}
\psfig{file=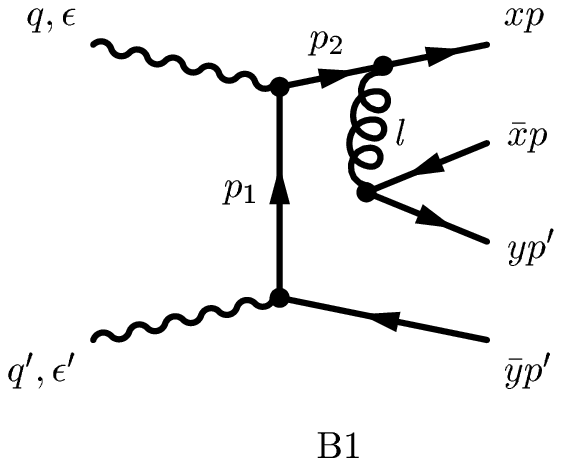,bb=200 640 380 790,width=5cm,angle=0}
\psfig{file=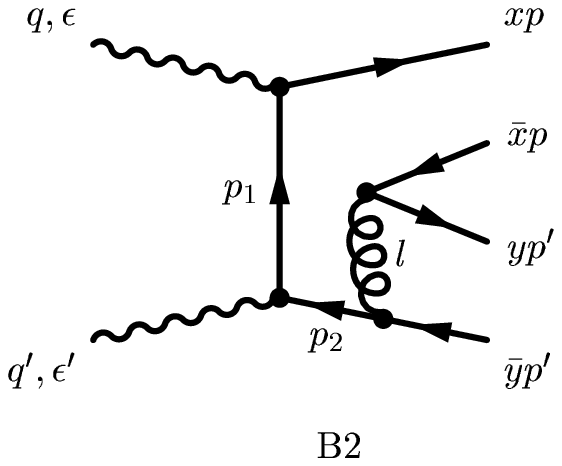,bb=200 640 380 790,width=5cm,angle=0}
\psfig{file=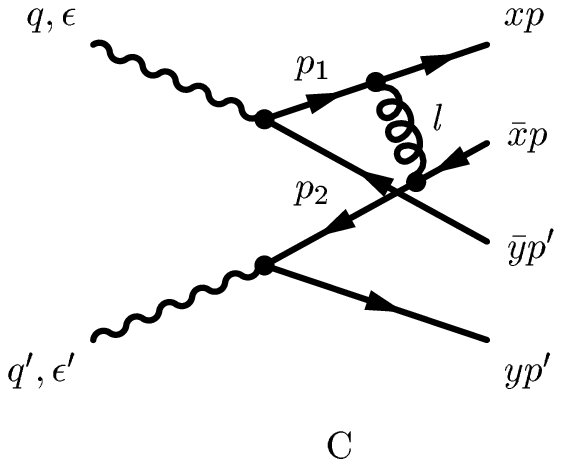,bb=200 640 380 790,width=5cm,angle=0}
\psfig{file=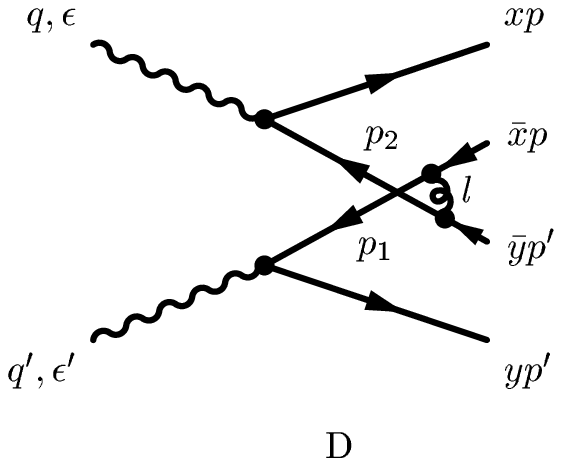,bb=200 640 380 790,width=5cm,angle=0}
\caption{The different classes of diagrams for the
         $\gamma^*\gamma\to\pi^+\pi^-$ amplitude in the
         hard-scattering approach.} \label{HSP diagrams}
\end{center}
\end{figure}

{\bf 3.}  In Ref.~\cite{Diehl1} the process
$\gamma^*\gamma\to\pi^+\pi^-$ was considered in the kinematical region
where the squared c.m.\ energy $s$ is much smaller than the photon
virtuality $Q^2$.  We will now turn to a perturbative investigation of
this process in the well-known standard hard-scattering picture
\cite{BrodskyLepage}, where we impose $s, -t, -u \gg \Lambda^2$ and
the additional constraint $Q^2 \gg s$.  The scattering amplitude is
given by a convolution of a hard-scattering amplitude $T_{H}$ and two
single-pion distribution amplitudes $\phi_\pi$ \cite{BrodskyLepage2}:
\begin{eqnarray}
 {\cal M}_{\lambda\lambda'}(s,t,Q^2)=\frac{f_\pi^2}{24}\, \int_0^1
 dx\, dy\,\phi_\pi(\bar{y},\mu_F)\,\phi_\pi(x,\mu_F)\,
   T_{H,\lambda\lambda'}(x,y,s,t,Q^2), \label{shsp}
\end{eqnarray}
with $f_\pi\simeq 133$~MeV, where the integral is over momentum
fractions $x,\, y$ carried by the quarks in their respective parent
hadrons (cf. Fig.~\ref{HSP diagrams}). We only have to take into
account the leading $q\bar{q}$ Fock state for each of the pions, since
higher Fock state contributions are suppressed by powers of
$\alpha_s/s$. In contrast to (\ref{helamp0}), the hard scattering in
(\ref{shsp}) thus involves only $u$ and $d$ quarks.

The single-pion distribution amplitudes embody soft physics, and their
asymptotic form is known from the evolution equation. The general
expansion of $\phi_\pi$ upon Gegenbauer polynomials is given by
\begin{eqnarray}
 \phi_\pi(x,\mu_F)= 6 x\bar{x}\,\sum_{n\; {\rm even}} a_n(\mu_F)\,
    C_n^{3/2}(2x-1), \label{gegenbauer expansion}
\end{eqnarray}
where the coefficients obey ERBL evolution
\cite{BrodskyLepage,EfremovRadyushkin}:
\begin{eqnarray}
 a_0=1,\quad
 a_n(\mu_F)=
  \bigg( \frac{\alpha_s(\mu_F)}{\alpha_s(\mu_0)} 
  \bigg)^{-\gamma_n/(2\beta_0)}
 a_n(\mu_0),
\end{eqnarray}
and $\mu_F$ is the factorization scale. We will comment later on the
question of choosing this scale in our process. A phenomenological
investigation of the data for the $\pi\gamma$-transition form factor
\cite{CLEO} shows that the shape of the single-pion distribution
amplitude is close to the asymptotic form already at low scale
\cite{Musatov}. This is consistent with other high-energy processes of
the pion (see for instance \cite{JakobKrollRaulfs,Feldmann:1999uf} and
references therein). Hence it is sufficient to consider small
deviations of the single-pion distribution amplitude from its asymptotic form
by keeping only a finite number of Gegenbauer coefficients in
(\ref{gegenbauer expansion}).

In order to extract the large $s$-limit of the two-pion distribution
amplitude $\phitwopi^q$, we proceed by explicitly calculating
(\ref{shsp}), expanding the result in powers of $s/Q^2$, and comparing
the leading term with expression (\ref{helamp0}).

The hard-scattering kernel $T_H$, which describes the process at
parton level, is given by 20 Feynman diagrams. These 20 diagrams can
be obtained by particle interchange from four basic diagrams,
representatives of which are shown in Fig.~\ref{HSP diagrams}.  Note
that for later use we have depicted two examples (B1, B2) for the
second class of Feynman graphs.

In light-cone gauge the leading contributions from the diagrams of
group B are independent of $Q$ and proportional to $1/s$. They turn
out to be the dominant ones, and will be discussed in detail later on.

We now investigate the contributions from the diagrams in group A. The
gluon propagator in light-cone gauge takes the form
\begin{eqnarray} 
 \frac{i\delta_{ab}}{l^2+i\epsilon} 
\left( -g^{\mu\nu}+
\frac{l^\mu q'^\nu+q'^\mu l^\nu}{l\cdot q'} \right) ,
  \label{gluon prop}
\end{eqnarray}
where $l$ is the momentum of the gluon and $q'$ the one of the real
photon. The $i\epsilon$ in the denominator of (\ref{gluon prop}) can
be dropped because the corresponding singularities only occur at the
end-points of the $x$- and $y$-integrations in (\ref{shsp}) and are
canceled by the end-point zeroes of the single-pion distribution
amplitudes (\ref{gegenbauer expansion}).  In Feynman gauge the
diagrams of group A are of order $1/s$. To this order, however, the
extra term in the light-cone gauge gluon propagator cancels the
contribution from the $g^{\mu\nu}$-term.  The remaining contributions
to the amplitude are of order $1/Q$ for longitudinal and $1/Q^2$ for
transverse polarization of the virtual photon, and hence suppressed
compared with diagrams B.  This could be expected on general grounds
from a similar reasoning as in deep inelastic scattering, the
$\pi\gamma$-transition form factor~\cite{BrodskyLepage}, or the
factorization proofs of exclusive reactions \cite{Collins}: a gluon
which couples between hard and soft parts of the process can be
eliminated by a gauge transformation to leading order in $1/Q$. In
Feynman gauge the gluon exchange of diagrams A does contribute to
leading order and precisely corresponds to the path ordered
exponential of gauge fields in the definition of the two-pion
distribution amplitude.

The contributions of the cat's ears diagrams C and D are suppressed by
powers of $1/Q^2$ for kinematical reasons: all three internal partons
have virtualities of the order $Q^2$.  In leading order, the
hard-scattering amplitude of these diagrams is given by
\begin{eqnarray} 
 T_{H}^{C+D}(x,y,t,u)\sim
    \frac{-2tu}{(u+yt)\,yt\,(\bar{x}u+yt)}
    +\frac{-2tu}{(u+\bar{y}t)\,\bar{y}t\,(xu+\bar{y}t)} 
    + (x\leftrightarrow \bar{y},\, t\leftrightarrow u) \,,
  \label{diagrams-CD}
\end{eqnarray}
where in general both $-t$ and $-u$ are of order $Q^2$ as we see in
eq.~(\ref{tu}). Hence these diagrams come at most with a factor
$1/Q^2$.

A more detailed argument is needed when $\zeta$ is so close to 1 or 0
that $-t$ or $-u$ is much smaller than $Q^2$, while still being large
compared with $\Lambda^2$. This is not excluded by the kinematical
limit (\ref{zeta constraint}). If in addition $x$ or $y$ is close to
its end-points the hard-scattering kernel in (\ref{diagrams-CD})
becomes much larger than $1/Q^2$. One can however readily see that
with the phase space suppression in the integrals over $x$ and $y$ and
the end-point zeroes of the single-pion distribution amplitudes the
corresponding contributions to the $\gamma^*\gamma$ amplitude
(\ref{shsp}) are again of order $1/Q^2$. The same holds for the
contributions from diagrams A.

Having identified those graphs that represent the leading order
contributions in perturbation theory, we now proceed by discussing the
graphs of group B in more detail and relating them to the scheme of
Ref.~\cite{Diehl1}.  The interpretation of the right-hand parts of
these diagrams in terms of a two-pion distribution amplitude requires
the virtualities of the quark $p_2$ and the gluon $l$ in Fig.~\ref{HSP
diagrams} to be much smaller than the dominant hard scale $Q^2$, since
the very essence of factorization is that, up to the logarithmic
effects of evolution, the two-pion distribution amplitude is
independent of $Q^2$.  This is indeed the case: explicitly we have
$p_2^2=ys$ and $l^2=\bar{x}ys$ in diagram B1 and $p_2^2=\bar{x}s$ and
$l^2=\bar{x}ys$ in diagram B2.  The physical picture is then the
following: a scattering of hardness $Q^2$ produces a quark and
antiquark, which on a scale $Q^2$ are quasi on-shell. In a second
step, with typical virtualities of order $s$, the quark or antiquark
radiates a gluon which in turn splits into a second
$q\bar{q}$-pair. Thirdly, the two quark-antiquark pairs rearrange to
hadronize into two pions. In the scheme represented in
Fig.~\ref{sketch} the second and third steps are described by the
two-pion distribution amplitude. We stress that this intuitive
physical picture only emerges in light-cone gauge, where the diagrams
of group A are subdominant.

In the diagrams of group B the photon can couple either to a $u$ or to
a $d$ quark.  For a distinct quark flavor $q$ there are two leading
order hard-scattering amplitudes, which we denote by $T_{H}^{q,B1}$
and $T_{H}^{q,B2}$, coming from diagrams B1 and B2, respectively.
Each of these two terms includes the contribution from the
corresponding graph obtained by permutation of the photon vertices.
In case of the photons coupling to the $u$ quark the hard-scattering
amplitude for diagrams B1 reads, up to corrections of ${\cal
O}(s/Q^2)$:
\begin{eqnarray}
 T_{H,\lambda\lambda'}^{u,B1}(x,y,s,t,u)= 
  - \frac{16\pi\alpha_s}{3}\, e^2 e_u^2 \,
   \delta_{\lambda\lambda'}\,
  \frac{2}{\bar{x} y s}\, \frac{u + (2y-1) t}{\bar{y} (u + y t)}\,
  \frac{(2-x) u + y t}{\bar{x} u + y t} \,,
    \label{th}
\end{eqnarray}
where $\alpha_s$ depends on a renormalization scale $\mu_R$. It
appears natural to take $\mu_R^2$ of order $s$, which is the scale of
the gluon virtuality in diagrams B.  By exchange of
$x\leftrightarrow\bar{y}$ and $t\leftrightarrow u$ in (\ref{th}) one
readily obtains the expression for $T_{H}^{u,B2}$.  Coupling the
photons to the $d$ quark results in the substitution
$x\rightarrow\bar{x}, \; y\rightarrow\bar{y}$ and $e_u^2\rightarrow
e_d^2$.  We see that to leading order the two incoming photons have
the same helicity and that, consequently, the virtual photon is
transversely polarized. Contributions from longitudinal polarization
of the virtual photon are suppressed by $1/Q$. Photons with opposite
helicities contribute with $1/Q^2$. This is in complete accordance
with the helicity selection rule appearing in (\ref{helamp0}).

In order to make a direct comparison with (\ref{helamp0}), we express
the amplitudes in terms of the variables $z$ and $\zeta$, defined in
(\ref{def z}) and (\ref{def zeta}). To this end we change integration
variables in (\ref{shsp}). In the case where the photons couple to the
$u$ quark we set
\begin{eqnarray} 
  \bar{z} = \bar{\zeta}\, \bar{y} && {\rm for\quad B1},
  \label{substy}\\  
  z       = \zeta x\,             && {\rm for\quad B2}, 
  \label{substx}
\end{eqnarray}
while leaving unchanged the corresponding other variable, $x$ for
diagrams B1 and $y$ for diagrams B2. From the kinematical constraint
$0\le x,y \le 1$ it immediately follows that
\begin{eqnarray}
 \zeta \le z \le 1 &&{\rm for\quad B1}, \\
 0 \le z \le \zeta &&{\rm for\quad B2},
\end{eqnarray}
i.e.\ in the two-pion distribution amplitude diagrams B1 and B2
contribute to distinct intervals of $z$, separated by the point
$z=\zeta$. For the photons coupling to the $d$ quark we perform
analogous substitutions, keeping in mind that $z$ is the momentum
fraction of the quark in the two-pion distribution amplitude. After
convoluting with the two single-pion distribution amplitudes
$\phi_\pi$ and substituting as explained, we obtain the helicity
amplitude (\ref{shsp}) in the limit $\Lambda^2\ll s\ll Q^2$:
\begin{eqnarray} 
 {\cal M}_{\lambda\lambda'}(\zeta,s)= 
% -\frac{1}{2}\,
  \frac{1}{2}\,
  \delta_{\lambda\lambda'}\, 
  e^2 e_u^2 \,
  \frac{8\pi f_\pi^2}{9} \int_0^1 dz \,\frac{2z-1}{z\bar{z}}
  \Bigg\{&&\hspace{-7mm}\Theta(\zeta-z)\, \frac{\zeta}{\zeta-z} \, 
  \phi_\pi\bigg(\frac{z}{\zeta}\bigg) \; I(\bar{z},\bar{\zeta},s) 
\nonumber\\ 
  -&&\hspace{-7mm}\Theta(z-\zeta)\,\frac{\bar{\zeta}}{z-\zeta} \, 
  \phi_\pi\bigg(\frac{\bar{z}}{\bar{\zeta}}\bigg) \, I(z,\zeta,s)
  \Bigg\} \nonumber \\
&& \hspace{-15.5em} {}+ (e_u^2\rightarrow e_d^2, z\rightarrow \bar{z}) ,
                       \label{helamp}
\end{eqnarray}
where the first term in curly brackets originates from diagrams B2 and
the second one from B1, and the integral $I(z,\zeta,s)$ is defined by
\begin{eqnarray}
  \label{integral}
   I(z,\zeta,s)=\int_0^1 dx \, \frac{\alpha_s}{s} \, 
    \frac{z+\bar{x}\zeta}{z-x\zeta} \, \frac{\phi_\pi(x)}{\bar{x}}\,.
\end{eqnarray} 
We note that in the $e_d^2$\,-term of (\ref{helamp}) we have
implicitly used isospin symmetry of the single-pion distribution
amplitude, $\phi_\pi(x)=\phi_\pi(\bar{x})$, as is embodied in
(\ref{gegenbauer expansion}). Starting from (\ref{th}) and the other
contributions to $T_H$ it is straightforward to generalize our results
to mesons without this symmetry, such as kaons.  Since the factor in
front of the curly brackets in (\ref{helamp}) is odd under
$z\rightarrow\bar{z}$ it immediately follows that
\begin{eqnarray}
 \phitwopi^u(z,\zeta,s)=-\phitwopi^{d}(\bar{z},\zeta,s).
   \label{phi-ud}
\end{eqnarray}
According to our remark below eq.~(\ref{shsp}) it is also clear that
there is no strange quark contribution:
\begin{eqnarray}
 \phitwopi^s\equiv 0.
   \label{phi-s}
\end{eqnarray}
 
We are now able to read off the expression for the two-pion distribution 
amplitude by direct comparison with eq.~(\ref{helamp0}):
\begin{eqnarray}
 \phitwopi^u(z,\zeta,s)= \frac{8\pi f_\pi^2}{9} 
  \Bigg\{&&\hspace{-7mm}\Theta(\zeta-z)\, \frac{\zeta}{\zeta-z} \, 
  \phi_\pi\bigg(\frac{z}{\zeta}\bigg) \; I(\bar{z},\bar{\zeta},s) 
\nonumber\\
  -&&\hspace{-7mm}\Theta(z-\zeta)\, \frac{\bar{\zeta}}{z-\zeta} \,
   \phi_\pi\bigg(\frac{\bar{z}}{\bar{\zeta}}\bigg) \; 
   I(z,\zeta,s) \Bigg\}. \label{model DA}
\end{eqnarray}
In order to extract $\phitwopi$, we have made an identification of the
integrands, so it might be questionable whether such an identification
is unique. However, the right-hand parts of diagrams B describe the
two-pion distribution amplitude in any process at large invariant pion
mass.  In our calculation we have made an identification at fixed $z$,
without using the particular $z$-dependence of the hard-scattering
kernel for $\gamma^*\gamma\to q\bar{q}$. Thus our extraction of the
two-pion distribution amplitude is unambiguous.

We immediately see that the result (\ref{model DA}) satisfies the
condition (\ref{sym}) of charge conjugation invariance. To evaluate
the moments of (\ref{model DA}) we revert the variable substitutions
that lead from (\ref{th}) to (\ref{helamp}), i.e.\ for $z>\zeta$ we
set $\bar{z}=\bar{\zeta}\bar{y}$, and for $z<\zeta$ we first replace
$x\rightarrow\bar{y}$ and then set $z=\zeta x$.
%In combination with (\ref{phi-ud}) 
This gives
\begin{eqnarray}
\lefteqn{ \int_0^1 dz\, (2z-1)^{n}\, \phitwopi^u(z,\zeta,s) =
  \frac{8\pi f_\pi^2}{9} \int_0^1 dx\, dy\, \frac{\alpha_s}{s}\,
  \frac{\phi_\pi(x)}{\bar{x}}\, \frac{\phi_\pi(\bar{y})}{y} } 
\nonumber \\
&& \left[  \zeta (2x\zeta-1)^{n}
           - \bar{\zeta} (1-2\bar{y}\bar{\zeta})^{n} 
           - 2 \zeta \bar{\zeta} 
            \sum_{\scriptstyle i=1 \atop \scriptstyle {\rm odd}}^{n}
          \left( \begin{array}{c} n \\ i \end{array} \right)
          (x\zeta - \bar{y}\bar{\zeta} )^{n-i} \
          (1-x\zeta-\bar{y}\bar{\zeta})^{i-1} 
   \right] . \hspace{2em}
\end{eqnarray}
We see that indeed the $(n+1)$-moment of the two-pion distribution
amplitude is a polynomial in $\zeta$ of degree $n+1$ at
most.\footnote{For this it is necessary to make a suitable choice of
renormalization and factorization scales. Namely, they are only
allowed to depend on $\zeta$ via $x$ and $y$ in (\ref{shsp}), and this
dependence must be chosen the same for diagrams B1 and B2.} For the
first moment we explicitly have
\begin{equation}
\int_0^1 dz\, \phitwopi^u(z,\zeta,s) = (2\zeta-1)\,
  \frac{8\pi f_\pi^2}{9} \int_0^1 dx\, dy\, \frac{\alpha_s}{s}\,
  \frac{\phi_\pi(x)}{\bar{x}}\, \frac{\phi_\pi(\bar{y})}{y} .
\end{equation}
In the term multiplying $(2\zeta-1)$ we recognize the expression of
$F_\pi(s)$ in the collinear hard-scattering approximation. Together
with (\ref{phi-ud}) and (\ref{phi-s}) our result thus fulfills the sum
rule~(\ref{sum-rule}).  
We note that within the collinear hard-scattering picture one also
finds that for large $s,-t,-u$ but $Q^2=0$ the helicity non-flip
amplitude $\cal{M}_{\lambda\lambda'}$ is proportional to the timelike
pion form factor \cite{BrodskyLepage2}.

{}From (\ref{integral}) we see that the two-pion distribution
amplitude (\ref{model DA}) falls off like $1/s$, up to logarithmic
corrections. Like for the pion form factor, such a power-law behavior
is a characteristic feature of the collinear hard-scattering
mechanism.

Note also that (\ref{model DA}) is a purely real function within our
phase convention, where the single-pion distribution amplitude is
real. This is because the Born level hard-scattering diagrams B do not
have $s$-channel cuts. These only occur at higher orders in
$\alpha_s$, for instance in the self-energy corrections for the
gluon. In the large-$s$ limit the ratio of imaginary and real part of
the two-pion distribution amplitude is thus suppressed by $\alpha_s$.

\begin{figure}[ht]
\begin{center}
\psfig{file=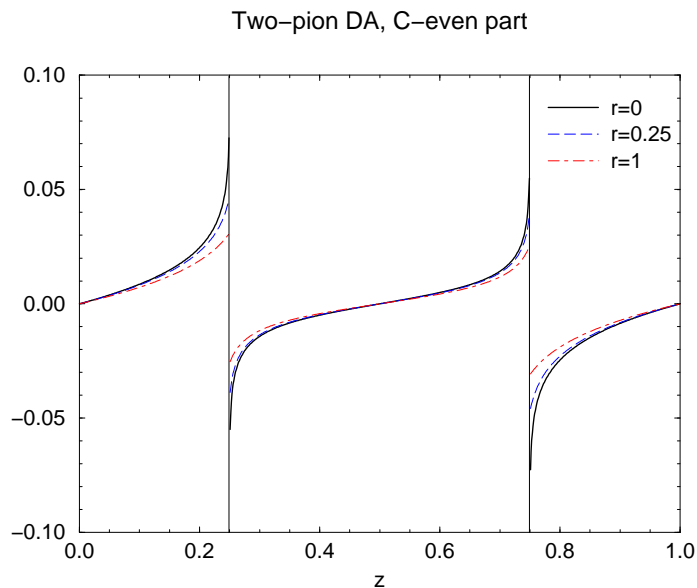,bb=30 80 560 670,width=8cm,angle=-90}
\caption{The $C$-even part of the two-pion distribution amplitude
         (\protect\ref{as twopida}) at $\zeta=0.25$ and $s=20$ GeV$^2$
         with various values of the cutoff parameter $r$ described in
         the text. The renormalization scale in $\alpha_s$ is chosen
         to be $s/4$.}  \label{c-even}
\end{center}
\end{figure}

\begin{figure}[ht]
\begin{center}
\psfig{file=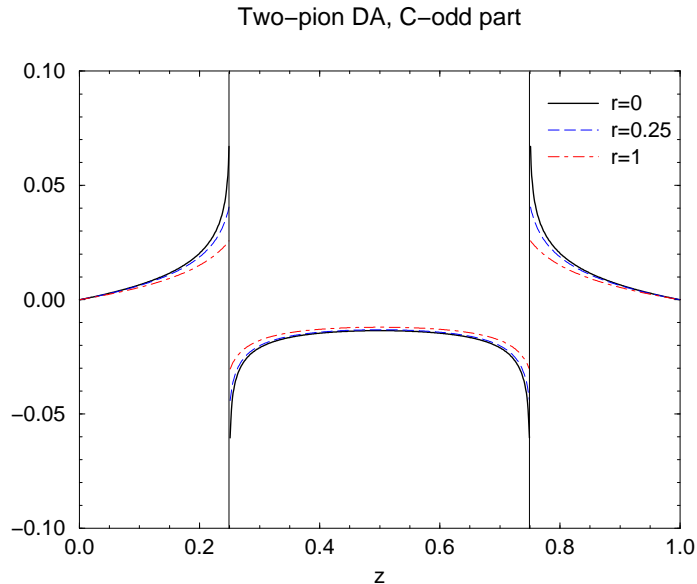,bb=30 80 560 670,width=8cm,angle=-90}
\caption{The $C$-odd part of the two-pion distribution amplitude
         (\protect\ref{as twopida}). Parameters are as in
         Fig. \ref{c-even}.}  \label{c-odd}
\end{center}
\end{figure}

Restricting ourselves to the asymptotic form of the single-pion
distribution amplitude as an example,
\begin{eqnarray}
  \phi_\pi(x)=\phi_{\rm AS}(x)=6x\bar{x} , 
\label{asy}
\end{eqnarray}
and 
taking an $x$-independent renormalization scale, the
two-pion distribution amplitude takes the following form: 
\begin{equation} 
 \phitwopi^u(z,\zeta,s)=-32\pi f_\pi^2\, \frac{\alpha_s}{s}
   \Bigg\{\Theta(\zeta-z)\,
   \bigg[\frac{z}{2\zeta}+\frac{z\bar{z}}{\zeta\bar{\zeta}}
   \ln\frac{\zeta-z}{\bar{z}}\bigg]-\Theta(z-\zeta)\,
   \bigg[\frac{\bar{z}}{2\bar{\zeta}}
   +\frac{z\bar{z}}{\zeta\bar{\zeta}}\ln\frac{z-\zeta}{z}\bigg]\Bigg\}.
\label{as twopida}
\end{equation}
It is useful to introduce the $C$-even and $C$-odd parts of
$\phitwopi^q$,
\begin{equation}
\phitwopi^{q, \pm}(z,\zeta,s) = \frac{1}{2} \left[ 
   \phitwopi^q(z,\zeta,s) \pm \phitwopi^q(z,\bar{\zeta},s) \right] ,
\end{equation} 
which project on the $\pi^+\pi^-$ system in states of definite
$C$-parity.  By virtue of (\ref{sym}) the $C$-even part
$\phitwopi^{q,+}(z,\zeta,s)$ is odd under $z\leftrightarrow \bar{z}$,
while $\phitwopi^{q,-}(z,\zeta,s)$ is even under this
exchange. Figures~\ref{c-even} and \ref{c-odd} show the $C$-even and
$C$-odd parts of the two-pion distribution amplitude~(\ref{as
twopida}).

The result (\ref{as twopida}) can easily be extended to more
complicated single-pion distribution amplitudes. Using standard
integrals it is possible to find the analytical form of the two-pion
distribution amplitude including an arbitrary number of terms in the
expansion (\ref{gegenbauer expansion}).  For the $a_2$-term for
instance the integral (\ref{integral}) reads
\begin{eqnarray}
  I(z,\zeta,s)= 36\, a_2 \, \frac{\alpha_s}{s} \,
    \bigg[-\frac{1}{12}
    +5\bigg(\frac{z}{2\zeta}-\frac{z^2}{\zeta^2}\bigg)
    -\bigg(\frac{z}{\zeta}-5\frac{z^2}{\zeta^2}
    +5\frac{z^3}{\zeta^3}\bigg)\ln\frac{z-\zeta}{z}\bigg]. 
    \label{non as}
\end{eqnarray} 

We must now comment on the logarithmic singularities at $z=\zeta$ of
the two-pion distribution amplitude (\ref{model DA}), which are
manifest in (\ref{as twopida}) and (\ref{non as}) and due to the fact
that at that point the integral $I(z,\zeta,s)$ in (\ref{integral}) is
divergent. They can be traced back to the singularity of $l\cdot q'$
in the light-cone gauge gluon propagator (\ref{gluon prop}) and show
up as the denominator $\bar{x} u + y t$ in the hard-scattering
amplitude $T_H^{u,B1}$ of eq.~(\ref{th}). This denominator cancels if
one adds to (\ref{th}) the contribution from diagrams B2, hence no
divergence is encountered in the $\gamma^*\gamma$ amplitude given by
(\ref{shsp}). We have however made different variable substitutions in
$T_H^{u,B1}$ and $T_H^{u,B2}$ in order to extract the two-pion
distribution amplitude, and after this the cancellation no longer
takes place.

Physically, the point $z=\zeta$ corresponds to the situation where one
of the intrinsic quarks in a pion carries all of the pion's momentum
whereas the other quark has zero four-momentum, cf.\
eqs.~(\ref{substy}) and (\ref{substx}). Of course, such a situation is
unphysical and due to the limit of massless partons in combination
with the collinear approximation employed when calculating the
hard-scattering amplitude. The singularity in $\phitwopi(z,\zeta,s)$
is a particular manifestation of the end-point singularities one
encounters in the collinear hard-scattering approach, a problem that
has been hotly debated for years \cite{isg}. In many applications,
such as the $\gamma^*\gamma$ amplitude ${\cal M}_{\lambda\lambda'}$,
these singularities are canceled by the end-point zeroes of the
distribution amplitudes $\phi_\pi$, and the result is finite. Note,
however, that if $s$ is not large enough this result is strongly
sensitive to the end-point regions, where the collinear approximation
breaks down and perturbative QCD cannot be reliably used because the
particle virtualities are small.  In the case of the two-pion
distribution amplitude the situation is even worse, since for
$z=\zeta$ not all end-point singularities are canceled. For $z$ close
to $\zeta$ the collinear hard-scattering approach can thus not be
applied to the hadronic matrix element $\phitwopi(z,\zeta,s)$ itself.

In order to understand this point on a more quantitative level, let us
introduce for the moment a regularization for the integral
(\ref{integral}) that cuts off the region where the longitudinal
parton momenta are soft, i.e.\ where $x$ or $\bar x$ are smaller than
$\Lambda^2/s$. Thus we replace the integration limits 0 and 1 in
(\ref{integral}) by $r\,\Lambda^2/s$ and $1-r\,\Lambda^2/s$,
respectively, where $r$ is a parameter that can be varied in order to
investigate the sensitivity to the cutoff.  In the limit $z \to \zeta$
the integral generates a logarithmic singularity $\propto \ln
(\Lambda^2/s)$ which cancels the logarithm in the running coupling
$\alpha_s$.  One is thus left with an expression at leading-power
level $1/s$ that formally is not of order $\alpha_s$.  This should be
taken as an indication that in the limit considered the two-pion
distribution amplitude (\ref{model DA}) is still sensitive to soft
physics which cannot be accounted for in the standard hard-scattering
picture.  As already explained, in the $\gamma^*\gamma$ amplitude the
logarithmic singularity drops out after integration over $z$ in
(\ref{helamp}), and our cutoff procedure only yields true power
corrections of order $\Lambda^2/s$, 
which go beyond the accuracy of the hard-scattering approximation and
can be neglected for sufficiently large $s$. Setting $\Lambda=1$~GeV,
$r=1$, and taking the asymptotic single-pion distribution amplitude
(\ref{asy}) we find for example at $s=20$ GeV$^2$ that 
${\cal M}_{\lambda\lambda'}$
decreases by 10\% compared with the result for no cutoff. A similar
suppression of the amplitude arises when the endpoint regions of the
$z$-integration in (\ref{helamp}) are cut off.

In Figs.~\ref{c-even} and \ref{c-odd} we show the results of our
cutoff regularization for the two-pion distribution amplitude, with
$\Lambda=1$~GeV and different values of $r$. We see that the results
are sensitive to soft physics around $z=\zeta$, and become stable when
$z$ is sufficiently far away from $\zeta$. From our discussion above
it is natural to expect that the hard-scattering calculation
becomes reliable when $|z-\zeta|$ is large compared with
$\Lambda^2/s$.

Let us finally comment on the question of factorization scales.  At
which scale the single-pion distribution amplitudes in (\ref{helamp})
and (\ref{integral}) should be taken is a priori not clear, since
there are two large scales $s$ and $Q^2$ present in the
hard-scattering diagrams. Likewise there is the question of the
factorization scale in the two-pion distribution amplitude we have
identified. The answer would require an analysis of the
$\alpha_s$-corrections to the diagrams we have calculated here. Notice
that at that level the two pions can not only originate from
$q\bar{q}$ but also from a gluon pair, and that the ERBL evolution of
the two-pion distribution amplitude for quarks mixes with the one for
gluons~\cite{Diehl1}. One does therefore not expect a simple relation
between the factorization scales of $\phi_\pi$ and of $\phitwopi$ in
(\ref{model DA}). We also note that our results in Figs.~\ref{c-even}
and \ref{c-odd}, obtained with the asymptotic expression of
$\phi_\pi(x)$, are far away from the asymptotic form of the two-pion
distribution amplitude, whose $z$-dependence is given by $\phitwopi^{+}
\sim z(1-z) (2z-1)$ and $\phitwopi^{-} \sim z(1-z)$. \\

{\bf 4.}  To summarize: We have shown how to obtain the two-pion
light-cone distribution amplitude in the perturbative limit, where
both squared c.m.\ energy $s$ and photon virtuality $Q^2$ are large
such that $\Lambda^2\ll s\ll Q^2$, where $\Lambda$ is of the order of
1 GeV.  Working in the collinear hard-scattering picture we have
identified the diagrams with handbag topology as the leading Feynman
graphs and expressed the two-pion distribution amplitude in terms of
the ordinary single-pion distribution amplitudes. Using light-cone
gauge, we have explicitly checked that a gluon attached to the hard
part of these diagrams leads to a suppression by a factor $1/Q$. This
is in accordance with general theory, where it is well known that
contributions from transverse gluons are power suppressed, whereas
scalar and longitudinal gluons can be eliminated by choosing an
appropriate gauge.  In this approach we have been able to reproduce
the scaling behavior and helicity selection rule of the process.  We
have verified the charge conjugation relation and the polynomiality
condition for the two-pion distribution amplitude, and that its first
moment equals the timelike pion form factor. Like the latter, our
two-pion distribution amplitude falls like $1/s$, and to leading order
in $\alpha_s$ is purely real.

Using expression (\ref{as twopida}) for the two-pion 
distribution amplitude we can write the amplitude for the 
process under consideration in the following simple form:
\begin{eqnarray}
{\cal M}_{\lambda\lambda'}(\zeta,s)
   = \delta_{\lambda\lambda'}\, e^2 ( e_u^2+e_d^2 )\, 
     \frac{8\pi f_\pi^2\, \alpha_s}{s}\,
 \bigg[\frac{1}{\zeta}\ln\frac{1}{\bar{\zeta}}+\frac{1}{\bar{\zeta}}
 \ln\frac{1}{\zeta}\bigg] ,
% \frac{d\sigma^{\gamma^*\gamma\to\pi^+\pi^-}}{d\zeta}
%   =\frac{\pi f_\pi^4\alpha_s^2}{s^3}\, 
% ( e_u^2+e_d^2 )^2\, 
% \bigg[\frac{1}{\zeta}\ln\frac{1}{\bar{\zeta}}+\frac{1}{\bar{\zeta}}
% \ln\frac{1}{\zeta}\bigg]^2,
\end{eqnarray}
where $\zeta$ must be sufficiently far from 0 and 1 to ensure $-t, -u
\gg \Lambda^2$.

We are aware that, at this stage, the investigations presented in this
work are above all theoretically motivated, since it will be difficult
to obtain experimental data for the process
$\gamma^*\gamma\to\pi^+\pi^-$ in the kinematical region we have
considered. However, we have seen how two different types of
factorization merge in the region where they are both applicable, and
know the form which any model two-pion distribution amplitude, for
instance the one proposed in \cite{Polyakov2}, has to approach at
large $s$. \\

{\bf Acknowledgments:} This work is partially supported by the U.S.\
Department of Energy, contract DE-AC03-76SF00515, and M.D.\
acknowledges support by the Alexander von Humboldt Foundation.
C.V. thanks the Deutsche Forschungsgemeinschaft for a graduate grant
and furthermore M. Polyakov and C. Weiss for useful discussions.

\end{document}